\documentclass[twocolumn,amsmath,amssymb]{revtex4-1}
\usepackage{graphicx} 
\usepackage{dcolumn} 
\usepackage{bm} 
\usepackage{braket} 
\usepackage[usenames,dvipsnames]{color}
\usepackage[
colorlinks=true, citecolor=blue, urlcolor=blue, linkcolor=blue,
setpagesize=false, bookmarks=false
]{hyperref}
\usepackage[english]{babel}

\begin{document}


\title{
Second-order magnetic responses in quantum magnets: \\ 
Magnetization under ac magnetic fields
}
\author{Tatsuya Kaneko$^{1}$, Yuta Murakami$^2$, Shintaro Takayoshi$^3$, and Andrew J. Millis$^{4,5}$}
\affiliation{
$^1$Computational Quantum Matter Research Team, RIKEN Center for Emergent Matter Science (CEMS), Wako, Saitama 351-0198, Japan\\
$^2$Department of Physics, Tokyo Institute of Technology, Meguro, Tokyo 152-8551, Japan\\
$^3$Department of Physics, Konan University, Kobe, Hyogo 658-8501, Japan\\
$^4$Department of Physics, Columbia University, New York, New York 10027, USA \\
$^5$Center for Computational Quantum Physics, Flatiron Institute, New York, New York 10010, USA
}

\date{\today}


\begin{abstract}
We investigate second-order magnetic responses of quantum magnets against ac magnetic fields. 
We focus on the case where the $z$ component of the spin is conserved in the unperturbed Hamiltonian and the driving field is applied in the $xy$ plane. 
We find that linearly polarized driving fields induce a second-harmonic response, while circularly polarized fields generate only a zero-frequency response, leading to a magnetization with a direction determined by the helicity. 
Employing an unbiased numerical method, we demonstrate the nonlinear magnetic effect driven by the circularly polarized field in the XXZ model 
and show that the magnitude of the magnetization can be predicted by the dynamical spin structure factor in the linear response regime. 
\end{abstract}

\maketitle


\section{Introduction}

Nonlinear responses are important phenomena for probing and controlling quantum states of matter under strong electromagnetic fields~\cite{ma2021}.  
Nonlinear responses of charge degrees of freedom to electric fields have been extensively investigated~\cite{orenstein2021,ghimire2019}. 
Recent work along these lines includes high-harmonic generation (HHG) realized in solids~\cite{ghimire2011,schubert2014,hohenleutner2015,vampa2015,liu2017,kaneshima2018,yoshikawa2019} and bulk photovoltaic effect in noncentrosymmetric materials~\cite{young2012,tan2016,cook2017,nakamura2017,osterhoudt2019,sotome2019,akamatsu2021}. 
These nonlinear phenomena are closely related to electronic band structures and collective excitations~\cite{sipe2000,vampa2014,tsuji2015,morimoto2016,ahn2020,murakami2021,kaneko2021,tanabe2021}.

In analogy to electronic charge responses, nonlinear effects driven by magnetic fields in quantum magnets are directly related to structures of spin excitations. 
Although the energy scale of spin excitations ($\sim J$ exchange coupling) is much lower than a charge gap in a typical magnetic insulator,  the development of the terahertz (THz) laser technique opens a pathway to address nonlinear magnetic effects associated with low-energy magnetic excitations~\cite{mukai2016,lu2017}. 
In this context, several nonlinear magnetic phenomena in the THz regime have been proposed theoretically. 
For example, magnetic HHG under linearly polarized fields~\cite{takayoshi2019,ikeda2019} and magnetization induced by circularly polarized fields~\cite{takayoshi2014_1,takayoshi2014_2} have been demonstrated numerically in driven quantum spin systems. 
While these numerical studies address the higher-order effects nonperturbatively, it is important to formulate the nonlinear phenomena in the perturbative regime, where the connection to equilibrium formulas allows for physical insight, which is most relevant to experiments. 

In this paper, we investigate second-order magnetic responses in quantum magnets, where the general structure of the theory can be elucidated. 
In particular, assuming that the $z$ component of spin is conserved in the unperturbed system, we derive the magnetic responses to ac magnetic fields applied perpendicular to the $z$ axis. 
We find that while the linearly polarized field (with the drive frequency $\Omega$) can produce a $2\Omega$ component of the magnetization, this $2\Omega$ oscillation is absent when circularly polarized fields are applied. 
However, applied circularly polarized fields produce a zero-frequency component of the magnetization, whose direction depends on the helicity of the applied field.
Our results are consistent with the numerical studies of  Refs.~\cite{takayoshi2019} and \cite{takayoshi2014_1}.
Specifically,  when the total magnetization vanishes in equilibrium and the excited states by the spin raising and lowering are symmetric in the spectrum, the spin magnetization $m^{(2)}_z$ at the second order is expressed as 
\begin{align}
\frac{dm^{(2)}_z}{dt}
 = i \gamma^2  \chi^{+-}_{\rm s} (\bm{q}=0,\Omega) 
\left[ \bm{B}(\Omega) \times \bm{B}(-\Omega)\right]_z ,
\notag
\end{align}
where $\gamma$ is the gyromagnetic ratio, $\chi^{+-}_{\rm s} (\bm{q},\Omega)$ is the dynamical (transverse) spin structure factor at the momentum $\bm{q}$, and $\bm{B}(\Omega)$ is the in-plane magnetic field at $\Omega$. 
We can see that the magnetization direction is controlled by the helicity of the applied magnetic field, and the magnitude is predicted by the dynamical spin structure factor in the linear response regime.  
This nonlinear magnetic effect owing to low-energy spin excitations should be contrasted with the inverse Faraday effect in metals, which is described by the cross product of the electric field $\bm{E}(\Omega)\times \bm{E}(-\Omega)$~\cite{hertel2006},  because the inverse Faraday effect in metals is essentially caused by electronic orbital degrees of freedom~\cite{battiato2014,berritta2016}. 
We also demonstrate the nonlinear magnetic effect in a driven XXZ model employing the infinite time-evolving block decimation (iTEBD) method~\cite{vidal2007} and confirm the above relation numerically. 

The rest of this paper is organized as follows.  
In Sec.~\ref{secModel}, we introduce the spin model that we address. 
In Sec.~\ref{sec:NLME}, we derive the magnetization in the second order and discuss the polarization dependence.   
Then, we focus on the magnetization induced by the circularly polarized field.
In Sec.~\ref{sec:numerics}, we provide the results of the numerical demonstration.  
Discussions and summary are given in Sec.~\ref{sec:summary}.


\section{Model} \label{secModel}

We consider a system under a magnetic field; 
\begin{align}
&\hat{\mathcal{H}}(t) = \hat{\mathcal{H}}_{0}  - \hbar \gamma \bm{B}(t) \cdot \hat{\bm{S}} , 
\label{eq:Ham}
\end{align}
where $\hat{\mathcal{H}}_0$ is the Hamiltonian of the spin system and the second term is the Zeeman coupling between the total spin $\hat{\bm{S}}$ ($= \sum_{j} \hat{\bm{S}}_j$) and the external magnetic field $\bm{B}(t)$.   
$\hat{S}^{\nu}_j$ ($\nu = x,y,z$) is the spin operator at site $j$ and $\gamma = -g \mu_{\rm B} /\hbar$ is the gyromagnetic ratio, where $g$ is the $g$ factor, $\mu_{\rm B}$ is the Bohr magneton, and $\hbar$ is the Planck constant. 
In this paper, we assume $[ \hat{\mathcal{H}}_{0},\hat{S}^z] = 0$, i.e., the eigenstate $\ket{\psi_m}$ of $\hat{\mathcal{H}}_0$ satisfies $\hat{\mathcal{H}}_0\ket{\psi_m} = \hbar \omega_m\ket{\psi_m}$ and $\hat{S}^z  \ket{\psi_m}  = S^z_m\ket{\psi_m}$, where $\hbar \omega_m$ and $S^z_m$ are the eigenenergy and quantum number of $\hat{S}^z$, respectively. 
In this paper, we focus on the magnetization $M_z(t) = \braket{\hat{S}^z(t)}$ under the magnetic field $\bm{B}(t) = (B^x(t),B^y(t),0)$.
The magnetization in the ground state is $M^{(0)}_z = \braket{\psi_0| \hat{S}^z |\psi_0}=S^z_0$.

  
\section{Second-order magnetic effects} \label{sec:NLME}

\subsection{Magnetization} 

We derive the field-induced magnetization using the perturbation theory at zero temperature  (see details in Appendix~\ref{appPT}). 
Note that we do not assume $M^{(0)}_z = 0$ at this stage. 
The magnetization at the first order in $\bm{B}(t)$ vanishes [i.e., $M^{(1)}_z(t)=0$] because the perturbation $\hat{\mathcal{V}}(t)=- \hbar \gamma [ B^x(t) \hat{S}^x + B^y(t) \hat{S}^y]$ induces the spin flip ($\hat{S}^{\pm}$) and $\braket{\psi_0 |\hat{S}^{z} \hat{S}^{\pm} | \psi_0}=0$. 
Then, the lowest order of the field-induced magnetization is of the second order. 
Using the perturbative expansion (see Appendix~\ref{appPT}), the time-dependent magnetization at the second order $M_z^{(2)}(t) $ is given by 
\begin{widetext}
\begin{align}
M_z^{(2)}(t) 
&= \frac{\gamma^2}{4 } \int^{t}_{\! -\infty} \! dt_1 \int^{t}_{\! - \infty} \! dt_2  \sum_{\zeta=\pm} \sum_{m} \zeta |\braket{\psi_m^{\zeta} | \hat{S}^{\zeta}|\psi_0}|^2 e^{-i(\omega_m^{\zeta}-\omega_0)(t_1-t_2)}  \bm{B}(t_1)\cdot \bm{B}(t_2)  
\notag \\
&+\frac{\gamma^2}{4i } \int^{t}_{\! -\infty} \! dt_1 \int^{t}_{\! - \infty} \! dt_2  \sum_{\zeta=\pm} \sum_{m} |\braket{\psi_m^{\zeta} | \hat{S}^{\zeta}|\psi_0}|^2 e^{-i(\omega_m^{\zeta}-\omega_0)(t_1-t_2)} \left[ \bm{B}(t_1) \! \times \! \bm{B}(t_2) \right]_z  ,
\label{eq:Mzt_all}
\end{align}
where $\hbar \omega_0$ is the ground-state energy of $\ket{\psi_{0}}$ and $\hbar \omega^{\zeta}_m$  ($\zeta = \pm$) is the eigenenergy of $\ket{\psi_m^{\pm}}$, in which $\hat{S}^z  \ket{\psi_m^{\pm}} = (M_z^{(0)}\pm 1)  \ket{\psi_m^{\pm}}$. 
When the magnetic field $\bm{B}(t_1)= \sum_{\Omega_1} \bm{B}(\Omega_1)e^{-i\Omega_1 t_1}$ is applied adiabatically from $t_1=-\infty$, the magnetization $M_z^{(2)}(t)$ is given by  
\begin{align}
M_z^{(2)}(t) = 
&-\frac{\gamma^2}{4 } 
\sum_{\Omega_1,\Omega_2}
\sum_{\zeta=\pm} \sum_{m}
\frac{e^{-i( \Omega_1 + \Omega_2 ) t }}{\Omega_1 + \Omega_2 + 2 i 0^+} \! 
\left[
\frac{ \zeta |\braket{\psi_m^{\zeta} | \hat{S}^{\zeta}|\psi_0}|^2}{\Omega_1 +\omega_m^{\zeta}-\omega_0 + i0^+ }      
+ \frac{ \zeta |\braket{\psi_m^{\zeta} | \hat{S}^{\zeta}|\psi_0}|^2}{\Omega_2 -\omega_m^{\zeta}+\omega_0 + i0^+}      
\right]
\bm{B}(\Omega_1) \cdot \bm{B}(\Omega_2)
\notag \\
&- \frac{\gamma^2}{4 i } 
\sum_{\Omega_1, \Omega_2}
\sum_{\zeta=\pm} \sum_{m}
\frac{e^{-i( \Omega_1 + \Omega_2 ) t }}{\Omega_1 + \Omega_2 + 2 i 0^+} \!
\left[
\frac{ |\braket{\psi_m^{\zeta} | \hat{S}^{\zeta}|\psi_0}|^2}{\Omega_1 +\omega_m^{\zeta}-\omega_0 + i0^+ }      
+ \frac{  |\braket{\psi_m^{\zeta} | \hat{S}^{\zeta}|\psi_0}|^2}{\Omega_2 -\omega_m^{\zeta}+\omega_0 + i0^+}      
\right] 
\left[ \bm{B}(\Omega_1) \! \times \! \bm{B}(\Omega_2) \right]_z.
\label{eq:Mag_tdep}
\end{align} 
\end{widetext}

\subsection{Second-harmonic generation}  \label{sec:SHG}

The oscillation of the magnetization with $\omega = n \Omega$ ($\Omega$: driving frequency) corresponds to high-harmonic generation from magnetic dipoles~\cite{takayoshi2019}. 
When a monochromatic field $\bm{B}(t) = \bm{B}(\Omega) e^{-i\Omega t} + \bm{B}(-\Omega) e^{i\Omega t}$ is applied, the magnetization at the second order $M_z^{(2)}(t)$ in Eq.~(\ref{eq:Mag_tdep}) has the component of second-harmonic generation (SHG) at $\Omega_1 = \Omega_2 = \Omega$.
Since $\bm{B}(\Omega) \times  \bm{B}(\Omega)=0$, 
the Fourier component $M^{(2)}_z(\omega=\Omega_1+\Omega_2 = 2\Omega)$ is given by 
\begin{widetext}
\begin{align}
M_z^{(2)}(2\Omega) = 
&-\frac{\gamma^2}{4} 
 \sum_{\zeta=\pm} \sum_{m}
 \frac{\zeta}{2 (\Omega + i0^+)}  
\left[ 
 \frac{ |\braket{\psi_m^{\zeta} | \hat{S}^{\zeta}|\psi_0}|^2 }{\Omega - \omega_m^{\zeta} + \omega_0 + i0^+}  
 +\frac{ |\braket{\psi_m^{\zeta} | \hat{S}^{\zeta}|\psi_0}|^2 }{\Omega + \omega_m^{\zeta} - \omega_0 + i0^+}  
\right]      
\bm{B}(\Omega) \cdot \bm{B}(\Omega). 
\label{eq:MagSHG}
\end{align}
\end{widetext}
The above Eq.~(\ref{eq:MagSHG}) is written as 
\begin{align}
M_z^{(2)}(2\Omega) = 
\alpha_z(2\Omega;\Omega,\Omega)
\bm{B}(\Omega) \cdot \bm{B}(\Omega). 
\end{align}

\begin{figure}[t]
\begin{center}
\includegraphics[width=0.96\columnwidth]{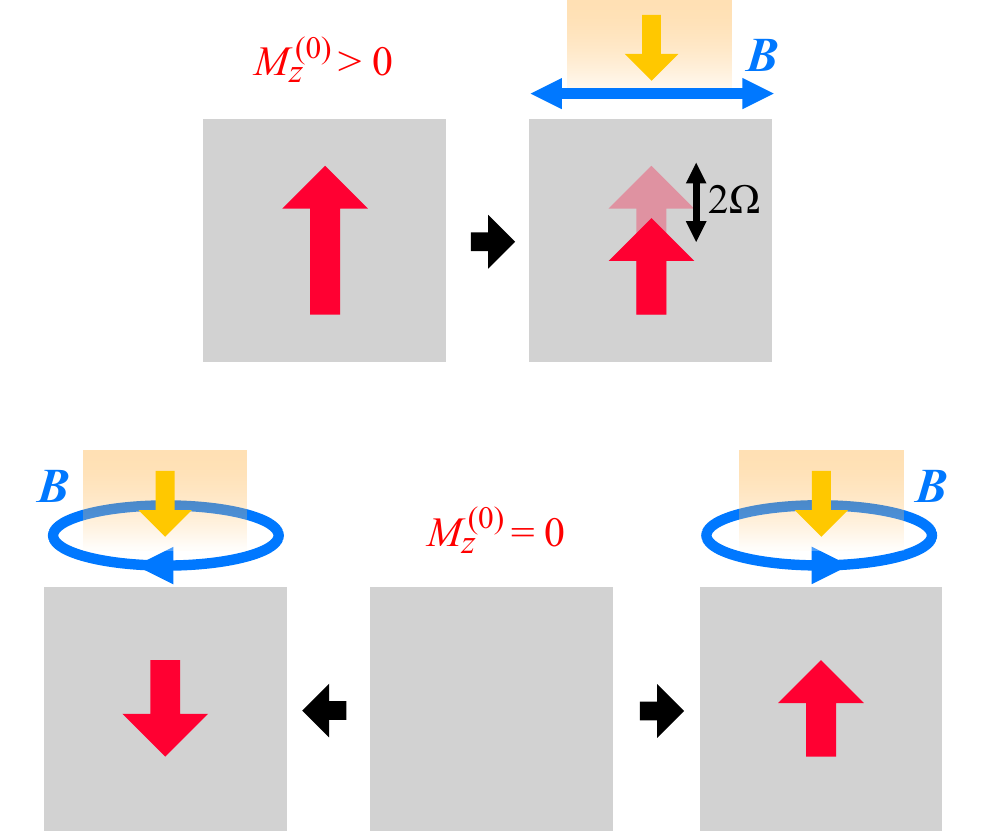}
\caption{
Schematic pictures of the second-order magnetic effects. 
Top panel: Magnetization under the linearly polarized magnetic field when $M^{(0)}_z>0$ in the ground state, where $M^{(2)}_z(2\Omega)$ is activated. 
Bottom panel: Helicity-dependent magnetization under the circularly polarized magnetic fields when $M^{(0)}_z=0$.  
}
\label{fig1}
\end{center}
\end{figure}

When a linearly polarized magnetic field $\bm{B}(t) = B_{\rm L} (\cos \Omega t, \cos \Omega t)$ [i.e., $\bm{B}(\Omega)= B_{\rm L}/2 (1,1)$] is applied, we find 
\begin{gather}
M_z^{(2)}(2\Omega) = \frac{1}{2}\alpha_z(2\Omega;\Omega,\Omega) B^2_{\rm L}, 
\label{eq:SZF_LP}
\end{gather}
and thus magnetic SHG is allowed.  
Note that $\alpha_z(2\Omega;\Omega,\Omega)=0$ when $|\braket{\psi^+_m | \hat{S}^+|\psi_0}|^2  =  |\braket{\psi^-_m | \hat{S}^-|\psi_0}|^2$ and $\omega_m^+ = \omega_m^-$, implying that $M_z^{(2)}(2\Omega)$ is nonzero only when the spin-flipped states by $\hat{S}^+$ and $\hat{S}^{-}$ are asymmetric.  
For example, $M_z^{(2)}(2\Omega)\ne 0$ when the ground state $\ket{\psi_0}$ has net magnetization $|M^{(0)}_z| > 0$~\cite{takayoshi2019}.   
A schematic picture of this effect is shown in the top panel of Fig.~\ref{fig1}. 

On the other hand, when a circularly polarized field $\bm{B}(t) = B_{\rm C} (\cos \Omega t, \pm \sin \Omega t)$ [i.e., $\bm{B}(\Omega)= B_{\rm C}/2 (1,\pm i )$]---where $\pm$ indicates the right- and left-handed circularly polarization---is applied, we find    
\begin{gather}
M_z^{(2)}(2\Omega) = 0, 
\end{gather}
because $\bm{B}(\Omega) \cdot \bm{B}(\Omega)=0$. 
Hence, in contrast to the response under the linearly polarized field, $M_z^{(2)}(2\Omega)$ is absent under the circularly polarized field. 

\subsection{Zero-frequency component}  \label{sec:ZFC}

The second-order magnetization $M_z^{(2)}(t)$ in Eq.~(\ref{eq:Mag_tdep}) also has a zero-frequency component at $\Omega_1 + \Omega_2 = 0$. 
Because $[\Omega_1 + \Omega_2 + 2i0^+]^{-1}$ diverges at $\Omega_1 + \Omega_2 =0$, Eq.~(\ref{eq:Mag_tdep}) is not a well-defined formula for describing the zero-frequency component.  
In order to get rid of the divergence arising from $\Omega_1 + \Omega_2 =0$, we consider the time derivative of $M_z^{(2)}(t)$. 
When $\bm{B}(t) = \bm{B}(\Omega) e^{-i\Omega t} + \bm{B}(-\Omega) e^{i\Omega t}$, the time derivative of $M_z^{(2)}(t)$ is given by 
\begin{align}
\frac{dM_z^{(2)}(t)}{dt} &= T^{(2)}_z(0) -i \sum_{n=\pm 1} 2 n \Omega M^{(2)}_z(2n\Omega)e^{-2in\Omega t} , 
\label{eq:DMagDt}
\end{align}
with the zero-frequency ($\omega = \Omega_1 + \Omega_2 = 0$) component 
\begin{widetext}
\begin{align}
T_z^{(2)}(0) & = 
\frac{\pi \gamma^2}{2} \sum_{\zeta=\pm} \sum_{m} \zeta |\braket{\psi_m^{\zeta} | \hat{S}^{\zeta}|\psi_0}|^2
\left[
\delta \left( \Omega -\omega_m^{\zeta}+\omega_0 \right)
+ \delta \left( \Omega +\omega_m^{\zeta}-\omega_0 \right) 
\right]
\bm{B}(\Omega) \cdot \bm{B}(-\Omega)
\notag \\
&+\frac{i \pi \gamma^2}{2}  \sum_{\zeta=\pm} \sum_{m}
|\braket{\psi_m^{\zeta} | \hat{S}^{\zeta}|\psi_0}|^2
\left[
\delta \left( \Omega -\omega_m^{\zeta}+\omega_0 \right)
- \delta \left( \Omega +\omega_m^{\zeta}-\omega_0 \right)
\right] 
\left[ \bm{B}(\Omega)  \times  \bm{B}(-\Omega) \right]_z.
\label{eq:MagZHG}
\end{align}
\end{widetext}
Equation~(\ref{eq:DMagDt}) implies that the magnetization grows linearly with $t$ when $T_z^{(2)}(0) \ne 0$. 
Because matrix elements of the total spin raising ($\zeta=+$) and lowering ($\zeta=-$) operators are involved, some properties of the first term in Eq.~(\ref{eq:MagZHG}) are linked to the equilibrium magnetization $M^{(0)}_z$. 
For example, when the spin is fully polarized with $M^{(0)}_z>0$, the first term gives the negative contribution because $ |\braket{\psi_m^{+} | \hat{S}^{+}|\psi_0}|^2 = 0$. 
On the other hand, this $\bm{B}(\Omega) \cdot \bm{B}(-\Omega)$ term vanishes if the $\zeta = + $ (raising) and $-$ (lowering) contributions are equivalent. 
This condition would require  $M^{(0)}_z=0$. 
Note that effects of relaxation are not taken into account in the above formula. 
The effects may be incorporated phenomenologically into Eq.~(\ref{eq:Mag_tdep}) by replacing $0^+$ with a relaxation factor $\Gamma$. 
In this case, the magnetization $M_z^{(2)}$ converges to a finite value of the order of $T_z^{(2)}(0) / \Gamma$.

Equation~(\ref{eq:MagZHG}) may be written as 
\begin{align}
T_z^{(2)}(0) & = 
\alpha_z'(0;\Omega,-\Omega)
\bm{B}(\Omega) \cdot \bm{B}(-\Omega)
\notag \\
&+i \beta_z' (0;\Omega,-\Omega) 
\left[ \bm{B}(\Omega) \times \bm{B}(-\Omega) \right]_z.
\label{eq:MagZHG_2}
\end{align}
When a linearly polarized magnetic field $\bm{B}(t) = B_{\rm L} (\cos \Omega t, \cos \Omega t)$ is applied, we find 
\begin{gather}
T_z^{(2)}(0) = \frac{1}{2}\alpha_z'(0;\Omega,-\Omega) B^2_{\rm L} . 
\label{eq:MZF_LP} 
\end{gather}
On the other hand, when a circularly polarized field $\bm{B}(t) = B_{\rm C} (\cos \Omega t, \pm \sin \Omega t)$ is applied, 
\begin{gather}
T_z^{(2)}(0)  =  \frac{1}{2} \left[ \alpha_z'(0;\Omega,-\Omega)  \pm  \beta_z'(0;\Omega,-\Omega) \right] B^2_{\rm C} .
\end{gather}
In both cases, the $\bm{B}(\Omega) \cdot \bm{B}(-\Omega)$ term can be nonzero.  
In contrast, the $\bm{B}(\Omega) \times \bm{B}(-\Omega)$ term can be nonzero only for a circularly polarized field. 
In this case, the magnetization exhibits helicity dependence.  
Hence, the nonlinear magnetic responses under the right- and left-handed circularly-polarized fields are asymmetric if $\alpha_z'(0;\Omega,-\Omega) \ne 0$. 
The bottom panel of Fig.~\ref{fig1} is a schematic picture of the effect when $M^{(0)}_z =0$ and $\alpha_z'(0;\Omega,-\Omega) = 0$.
As shown in Fig.~\ref{fig1}, we can manipulate the magnetization direction by the helicity of the magnetic field.

\subsection{Magnetization by circularly polarized fields}  \label{sec:MinCPF}

In the previous sections, we only assume the conservation of $S^z$ in the unbiased Hamiltonian $\hat{\mathcal{H}}_0$, and thus the expressions are general. 
Here, to see the nonlinear response can be connected to the dynamical spin structure factor, we focus on specific cases, where $|\braket{\psi^+_m | \hat{S}^+|\psi_0}|^2  =  |\braket{\psi^-_m | \hat{S}^-|\psi_0}|^2$ and $\omega_m^+ = \omega_m^-$ are satisfied in Eqs.~(\ref{eq:MagSHG}) and (\ref{eq:MagZHG}). 
These conditions may be realized, e.g., when $M^{(0)}_z=0$ and the Hamiltonian $\hat{\mathcal{H}}_0$ is invariant under the time-reversal operation (or $\pi$ rotation around the $y$ axis)
$\hat{S}^{\pm}_j \rightarrow - \hat{S}^{\mp}_j$ and $\hat{S}^{z}_j \rightarrow - \hat{S}^{z}_j$. 
When the above conditions are satisfied, $\alpha_z(2\Omega;\Omega,\Omega)=\alpha_z'(0;\Omega,-\Omega)=0$ and the response to a linearly polarized field vanishes [see Eqs.~(\ref{eq:SZF_LP}) and (\ref{eq:MZF_LP})]. 
However, even in this condition, the $\bm{B}(\Omega) \times \bm{B}(-\Omega)$ term in Eq.~(\ref{eq:MagZHG}) can be nonvanishing, implying that a magnetization $M^{(2)}_z(t)$ can be generated from $M^{(0)}_z=0$ by applying a circularly polarized field.  
Since $M_z^{(2)}(2\Omega)=0$ in Eq.~(\ref{eq:DMagDt}), the second-order magnetic response is described by 
\begin{widetext}
\begin{align}
\frac{d M_z^{(2)}}{dt} 
 =
i \pi \gamma^2  \sum_{m}  |\braket{\psi_m | \hat{S}^-|\psi_0}|^2
\left[ 
\delta \! \left( \Omega - \omega_m + \omega_0 \right) 
- \delta \! \left( \Omega + \omega_m - \omega_0 \right)
\right]  
\left[ \bm{B}(\Omega) \times \bm{B}(-\Omega)\right]_z ,
\label{eq:Tz_2nd}
 \end{align}
\end{widetext}
where we denote $\ket{\psi^{\pm}_m}$ and $ \omega_m^{\pm}$ by $\ket{\psi_m}$ and $ \omega_m$.  
 
Equation (\ref{eq:Tz_2nd}) is related to the commonly-used dynamical (transverse) spin structure factor 
\begin{align}
\chi^{+-}_{\rm s} (\bm{q},\Omega) 
&=  \pi \sum_{m} |\braket{\psi_m| \hat{S}^{-}_{\bm{q}}|\psi_0} |^2 \delta( \Omega-\omega_m+\omega_0), 
\end{align}
where $\hat{S}^{-}_{\bm{q}}=\frac{1}{\sqrt{N}}\sum_j \hat{S}^-_j e^{-i\bm{q}\cdot\bm{R}_j}$ [$N$: number of lattice sites] is the spin-flip operator in the momentum ($\bm{q}$) space.  
Using $\chi^{+-}_{\rm s}  (\bm{q},\Omega) $,  the magnetization per unit $m_z^{(2)}$($=M_z^{(2)}/N$) at $\Omega > 0$ is given by 
\begin{align}
\frac{dm_z^{(2)}}{dt}
 = i \gamma^2  \chi^{+-}_{\rm s} (\bm{q}=0,\Omega) 
\left[ \bm{B}(\Omega) \times \bm{B}(-\Omega)\right]_z .
\label{eq:NLME_CPL}
\end{align}
Hence, by introducing the structure factor $\chi^{+-}_{\rm s} (\bm{q},\Omega)$, we can describe the magnetization at the second order in the simple formula. 
Under a circularly polarized field $\bm{B}(\Omega)= B_{\rm C}/2 (1,\pm i )$, this magnetization exhibits the helicity ($\pm$) dependence 
\begin{align}
\frac{dm_z^{(2)}}{dt}
 =\pm \frac{1}{2}\gamma^2 \chi^{+-}_{\rm s}  (\bm{q}=0,\Omega)  B^2_{\rm C}. 
\label{eq:NLME_CPLpm}
\end{align}

For the magnetic effect described by Eq.~(\ref{eq:NLME_CPL}), the dynamical spin structure factor must be $\chi^{+-}_{\rm s} (\bm{q}=0,\Omega) \ne 0$. 
In other words, once we know the dynamical spin structure factor $\chi^{+-}_{\rm s} (\bm{q},\Omega)$ in the linear response regime, we can predict the main features of the magnetization at the second order. 
In the isotopic Heisenberg model [or spin-SU(2)-symmetric Hubbard model], $\chi^{+-}_{\rm s} (\bm{q}=0,\Omega) =0$ at $\Omega >0$ and no magnetization $m_z^{(2)}$ is induced. 
This implies that $\chi^{+-}_{\rm s} (\bm{q}=0,\Omega) \ne 0$ may arise from magnetic anisotropies, e.g., Ising anisotropy in the XXZ model (see Sec.~\ref{sec:numerics}) and the Dzyaloshinskii-Moriya interaction. 
As discussed in Appendix~\ref{appRF}, we can interpret this nonlinear magnetic effect in the rotating frame, where the system can be described by a static Hamiltonian~\cite{takayoshi2014_1,takayoshi2014_2}. 
 
Equation ~(\ref{eq:NLME_CPL}) is very similar to the formula of the circular photogalvanic effect (CPGE) in which a generated second-order photocurrent $J^{(2)}_{\mu}$ under an electric field $\bm{E}$ is described by $\frac{d }{dt} J^{(2)}_{\mu} = i \eta_{\mu}(\Omega) [\bm{E}(\Omega)\times \bm{E}(-\Omega)]_{\mu}$~\cite{sipe2000,juan2017}. 
While the magnetization and electric current are different, we may find similar time-dependent properties to the CPGE. 

\begin{figure}[t]
\begin{center}
\includegraphics[width=\columnwidth]{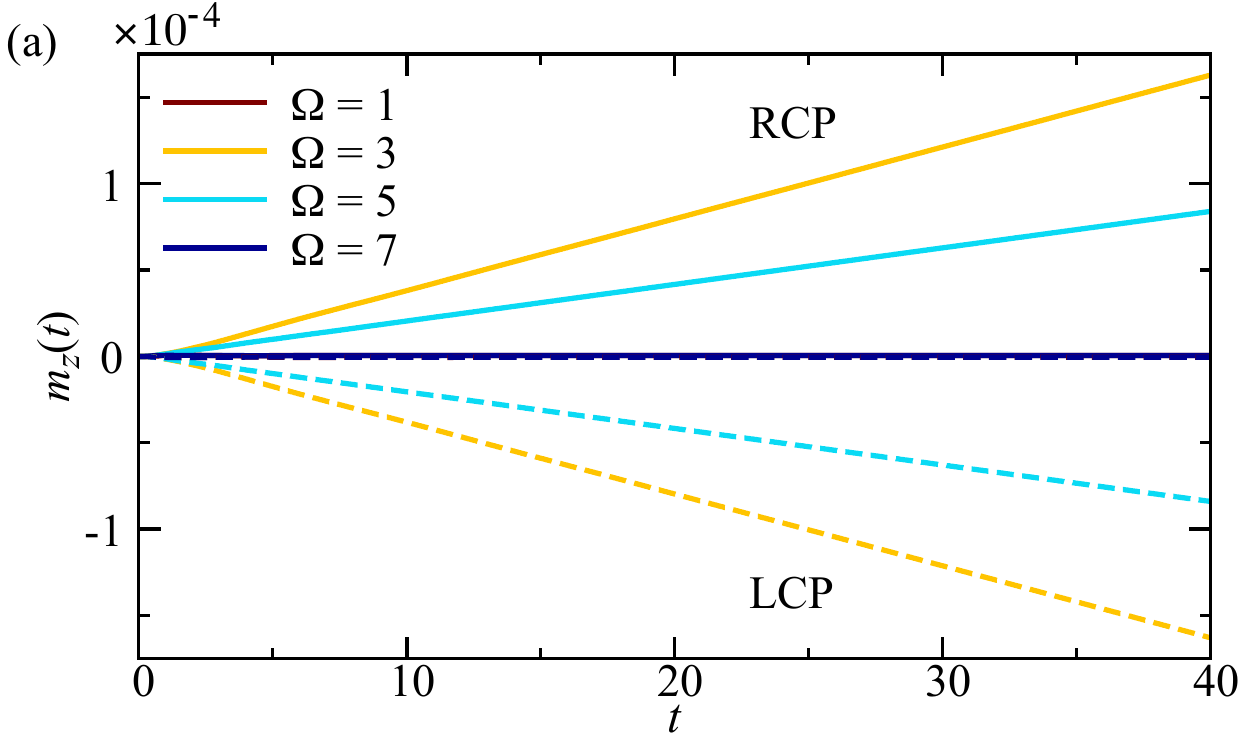}
\\
\includegraphics[width=\columnwidth]{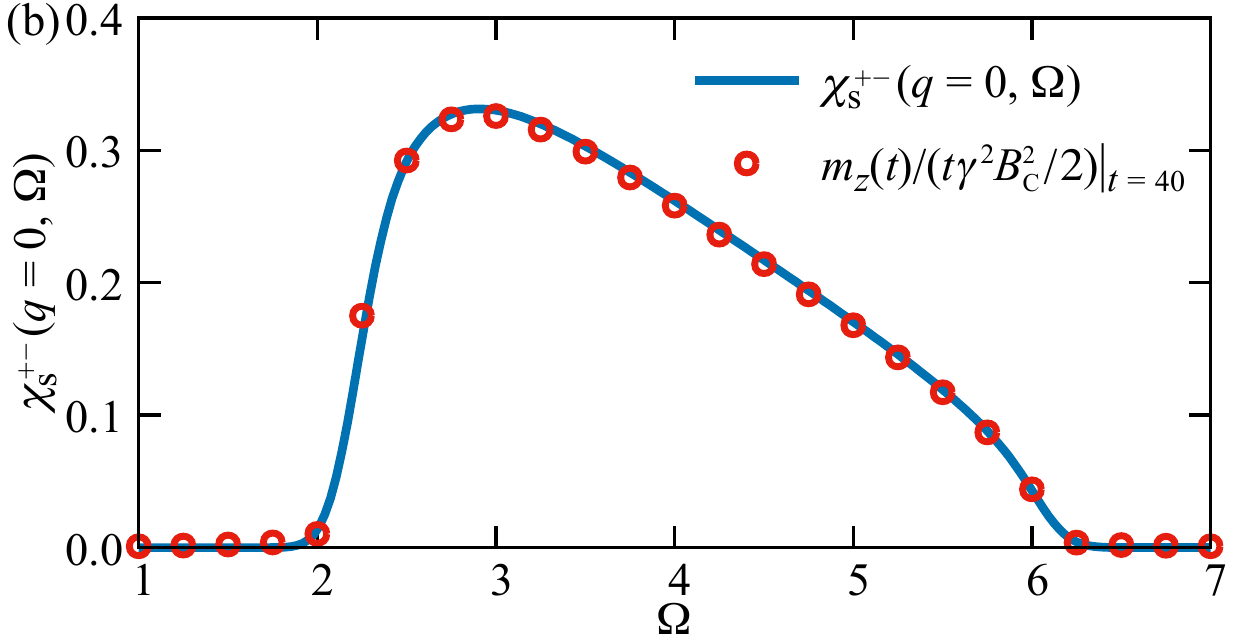}
\caption{(a) Time-dependent magnetization $m_z(t)$ under the magnetic field $\bm{B}(t) = B_{\rm C} (\cos \Omega t, \pm \sin \Omega t)$, where $J=1$, $\Delta=4$, and $\hbar \gamma B_{\rm C} = 0.005$ ($\hbar/J$ is  a unit of time). 
The solid and dotted lines indicate the magnetization under right-handed circularly polarized (RCP) and left-handed circularly polarized (LCP) fields, respectively. 
The data at $\Omega=1$ and 7 are overlapped around $m_z(t)\sim0$.   
(b) Comparison between the dynamical spin structure factor $\chi^{+-}_{\rm s}  (q=0,\Omega)$ (solid line) and the $\Omega$ dependence of the  magnetization normalized as $\left. m_z(t) / (t \gamma^2 B_{\rm C}^2/2)\right|_{t=40}$ (circles). 
}
\label{fig2}
\end{center}
\end{figure}


\section{Numerical demonstration} \label{sec:numerics}

Finally, we numerically demonstrate the nonlinear magnetic effect described by Eq.~(\ref{eq:NLME_CPL}) using the spin-1/2 XXZ model.  
As $\hat{\mathcal{H}}_0$ in Eq.~(\ref{eq:Ham}), the Hamiltonian of the one-dimensional XXZ model is 
\begin{align}
\hat{\mathcal{H}}^{\rm XXZ}_{0}  = J \sum_{j} \left[ \hat{S}_j^x \hat{S}_{j+1}^x+\hat{S}_j^y \hat{S}_{j+1}^y + \Delta \hat{S}_j^z \hat{S}_{j+1}^z\right], 
\end{align}
where $J>0$ is the antiferromagnetic exchange coupling and $\Delta$ is the magnetic anisotropy along the $z$ direction. 
Here, we set $J$ ($\hbar / J$) as a unit of energy (time). 
When $\Delta > 1$, the magnetic excitation in the XXZ chain is gapped and $\chi^{+-}_{\rm s} (q=0,\Omega)$ obtains the spectral weights above the gap.  
Thus, the second-order magnetic effect described by Eq.~(\ref{eq:NLME_CPL}) is anticipated. 
To demonstrate this effect, we employ the infinite time evolving block decimation (iTEBD)~\cite{vidal2007} and calculate the time dependence of $m_z(t)$ under the circularly polarized field $\bm{B}(t) = B_{\rm C} (\cos \Omega t, \pm \sin \Omega t)$.  

Figure~\ref{fig2}(a) shows the magnetization $m_z(t)$ at $\Delta=4$ in the XXZ model. 
Corresponding to $\chi^{+-}_{\rm s}  (q=0,\Omega)$ [see Fig.~\ref{fig2}(b)], the magnetization $m_z(t)$ is generated at $2 \lesssim \Omega \lesssim 6$. 
The sign of the magnetization is inverted by switching the helicity ($\pm$) of the magnetic field $\bm{B}(t)$. 
While the linear growth of the magnetization is expected at $t \gg 1$ (see Appendix~~\ref{appTP0}),
$m_z(t)$ already grows up linearly with time up to $t=40$. 
Since $i \gamma^2  \left[ \bm{B}(\Omega) \times \bm{B}(-\Omega)\right]_z = \gamma^2 B_{\rm C}^2/2$ in Eq.~(\ref{eq:NLME_CPL}), we plot the $\Omega$ dependence of the normalized magnetization $\left. m_z(t) / (t \gamma^2 B_{\rm C}^2/2)\right|_{t=40}$ in Fig.~\ref{fig2}(b). 
As plotted in Fig.~\ref{fig2}(b), the magnetization shows good agreement with  $\chi^{+-}_{\rm s}  (q=0,\Omega)$. 
Therefore, the second-order magnetic effect in the gapped phase of the XXZ model is actually described by Eq.~(\ref{eq:NLME_CPL}). 
While a similar numerical simulation has been performed in Ref.~\cite{takayoshi2014_1}, in our study, we formulate the nonlinear magnetic effect in a simple equation (\ref{eq:NLME_CPL}) and identify the relation with the low-energy magnetic excitation described by $\chi^{+-}_{\rm s} (q,\Omega)$.


\section{Summary and Discussion} \label{sec:summary}

In this paper, we have investigated the second-order magnetization perpendicular to the driving magnetic fields. 
We have derived that while $M_z(\omega=2\Omega)$ can be induced under the linearly polarized field, it is absent under the circularly polarized field. 
$M_z(\omega=0)$ can be induced by circularly polarized fields and exhibits helicity dependence.  
We have also discussed the specific case when the ground state has no net magnetization, where we have demonstrated the effect numerically in the driven XXZ model and have shown that the main features of the magnetization are determined by the dynamical spin structure factor $\chi^{+-}_{\rm s} (\bm{q}=0,\Omega)$. 

This second-order magnetic effect emerges in a quantum magnet with magnetic anisotropy. 
For example, BaCo$_2$V$_2$O$_8$ is described as an antiferromagnetic XXZ chain with $\Delta > 1$~\cite{kimura2007,grenier2015,faure2018}, where we may find a similar magnetic effect demonstrated in Fig.~\ref{fig2}. 
For $J \sim 3$~meV close to the value reported in BaCo$_2$V$_2$O$_8$~\cite{faure2018}, $\Omega=2$ and $\hbar \gamma B_{\rm C} =0.005$ in Fig.~\ref{fig2} correspond to $1.45$~THz and $0.13$~T~\footnote{$g\simeq 2$ and $\mu_{\rm B}=0.0579$~meV/T are used.}, respectively, which may be accessible in experiments. 
In the recently realized twisted WSe$_2$ that can be represented as a triangular lattice Hubbard model~\cite{wang2020,pan2020,zang2021}, the displacement field leads to a Dzyaloshinskii-Moriya-type anisotropic interaction in the effective Heisenberg model in the strong-coupling limit~\cite{pan2020,zang2021}. 
Because of the gapped magnetic excitation due to the anisotropic interaction, this moir\'e Hubbard system may also be a candidate for the host of the second-order magnetization. 
While in Fig.~\ref{fig2} we used a model that only has an excitation continuum, the relations we derived are exact regardless of the type of magnetic excitations. 
A magnetic collective mode, which gives a large response at a resonant excitation frequency in a dynamical spin correlation function, can be a good source for efficient nonlinear magnetic effects. 

In our study, effects of relaxation, which are present in any realistic systems (e.g., by spin-lattice relaxation), are not taken into account. 
When effects of relaxation are incorporated, the linear growth of the magnetization [e.g., in Fig.~\ref{fig2}(a)] is observed until the relaxation time $\tau$. 
The magnetization converges to a finite value in a steady state at $t \gg \tau$, where the magnitude of $m_z^{(2)}$ may be proportional to $\tau$. 

While we focus on the responses to the magnetic field component of a THz field, the electric field component is usually larger than the magnetic field component~\cite{ikeda2019}. 
Hence, if a spin-electric field coupling is crucial in a magnetic insulator, we might find a larger nonlinear magnetic response, which is useful for electromagnetic field manipulation of quantum materials.   
In order to address this issue, one needs to consider a coupling term between an electric field and a spin system via, e.g., spin-phonon or spin-orbit coupling. 
On the other hand, a recent technique using a split-ring resonator enables us to selectively enhance the strength of the THz magnetic field~\cite{mukai2014,kurihara2014,mukai2016}, which may also open a pathway to realize a large nonlinear magnetic effect.


\begin{acknowledgments}
This work was supported by Grants-in-Aid for Scientific Research from JSPS, KAKENHI Grants No.~JP18K13509 (T.K.), No.~JP20K14412, No.~JP20H05265, No.~JP21H05017 (Y.M.), and No.~JP21K03412 (S.T.) and JST CREST Grant No.~JPMJCR1901 (Y.M.) and No.~JPMJCR19T3 (S.T.).
T.K. was supported by the JSPS Overseas Research Fellowship. 
A.J.M. was supported in part by Programmable Quantum Materials, an Energy Frontier Research Center funded by the U.S. Department of Energy (DOE), Office of Science, Basic Energy Sciences (BES), under Award No. DE-SC0019443. 
The Flatiron Institute is a division of the Simons Foundation. 
\end{acknowledgments}


\appendix 

\section{Perturbation theory}  \label{appPT}

We employ the perturbation theory to derive a formula for the magnetization $M_z(t)$. 
With respect to the perturbation $\hat{\mathcal{V}}(t) = \hat{\mathcal{H}}(t) - \hat{\mathcal{H}}_{0}$, the wave function $\ket{\Psi(t)} = e^{-i\frac{\hat{\mathcal{H}}_0}{\hbar}t}\ket{\Psi_{\rm I}(t)}$ evolved from the ground state $\ket{\psi_0}$ is obtained via
\begin{align}
\ket{\Psi_{\rm I}(t)}  &= \ket{\psi_0} + \frac{1}{i \hbar} \int^{t}_{-\infty} dt_1 \, \hat{\mathcal{V}}_{\rm I}(t_1) \ket{\psi_0}
\label{eq:time_evo} \\
&+ \left( \frac{1}{i \hbar}\right)^2 \int^{t}_{-\infty} dt_1 \int^{t_1}_{-\infty} dt_2 \, \hat{\mathcal{V}}_{\rm I}(t_1)  \hat{\mathcal{V}}_{\rm I}(t_2) \ket{\psi_0} + \cdots ,
\notag 
\end{align}
where the subscript ${\rm I}$ indicates the interaction picture and $\hat{\mathcal{O}}_{\rm I}(t) =  e^{i\frac{\hat{\mathcal{H}}_0}{\hbar}t} \hat{\mathcal{O}}(t) e^{-i\frac{\hat{\mathcal{H}}_0}{\hbar}t}$. 
Assuming a transverse magnetic field, i.e., $B^z(t)=0$, in the Hamiltonian  (\ref{eq:Ham}), the perturbation term is given by
\begin{align}
\hat{\mathcal{V}}_{\rm I}(t) 
= - \hbar \gamma \left[ B^x(t) \hat{S}^x_{\rm I}(t) + B^y(t) \hat{S}^y_{\rm I}(t) \right].
\label{eq:perturb}
 \end{align}
Using the interaction picture, the magnetization is 
\begin{align}
M_z(t) = \braket{\Psi_{\rm I}(t)| \hat{S}^z_{\rm I}(t) |\Psi_{\rm I}(t)} . 
\end{align}

The magnetization in the ground state $\ket{\psi_0}$ is $M^{(0)}_z = \braket{\psi_0| \hat{S}^z |\psi_0}=S^z_0$. 
Although the magnetic field in Eq.~(\ref{eq:perturb}) is applied, the magnetization at the first order in $\bm{B}(t)$ vanishes, i.e., $M^{(1)}_z(t) = 0$, because $\hat{\mathcal{V}}_{\rm I}(t)$ induces the spin flip ($\hat{S}^{\pm}$) and $\braket{\psi_0 |\hat{S}^{z}_{\rm I}(t) \hat{\mathcal{V}}_{\rm I}(t') | \psi_0}=0$. 

Using Eq.~(\ref{eq:time_evo}), the magnetization at the second order $M_z^{(2)}(t)$ is given by
\begin{align}
M_z^{(2)}(t) &= \frac{1}{\hbar^2 } \int^{t}_{-\infty} \! dt_1 \int^{t}_{-\infty} \! dt_2 \braket{\psi_0|  \hat{\mathcal{V}}_{\rm I}(t_1) \hat{S}_{\rm I}^z(t) \hat{\mathcal{V}}_{\rm I}(t_2) |\psi_0} 
\notag \\
 &- \frac{1}{\hbar^2 } \int^{t}_{-\infty} \! dt_1 \int^{t_1}_{-\infty} \! dt_2 \braket{\psi_0|  \hat{S}_{\rm I}^z(t)  \hat{\mathcal{V}}_{\rm I}(t_1) \hat{\mathcal{V}}_{\rm I}(t_2) |\psi_0} 
 \notag \\
 &- \frac{1}{\hbar^2 } \int^{t}_{-\infty} \! dt_1 \int^{t_1}_{-\infty} \! dt_2 \braket{\psi_0|  \hat{\mathcal{V}}_{\rm I}(t_2) \hat{\mathcal{V}}_{\rm I}(t_1)\hat{S}_{\rm I}^z(t)  |\psi_0} . 
\label{eq:Mz_2nd}
\end{align}
Because $\hat{\mathcal{V}}_{\rm I}(t)$ is comprised of the spin-flip operators $\hat{S}^{\pm}$, we introduce the intermediate eigenstate $\ket{\psi_m^{\pm}}$ in which $S^z_m = S^z_0 \pm 1$. 
Using $\hat{S}^z  \ket{\psi_m^{\pm}} = (M_z^{(0)}\pm 1)  \ket{\psi_m^{\pm}}$, the integrand of the first term in Eq.~(\ref{eq:Mz_2nd}) is given by 
\begin{align}
& \braket{\psi_0|  \hat{\mathcal{V}}_{\rm I}(t_1) \hat{S}_{\rm I}^z(t) \hat{\mathcal{V}}_{\rm I}(t_2) |\psi_0} 
\notag \\
 &=\sum_{\zeta=\pm} \sum_{m}\left( M^{(0)}_z + \zeta \right)\braket{\psi_0|  \hat{\mathcal{V}}_{\rm I}(t_1) | \psi_m^{\zeta}} \braket{\psi_m^{\zeta} |  \hat{\mathcal{V}}_{\rm I}(t_2)  |\psi_0} .
 \label{eq:Mz_2nd_inte1st}
\end{align}
The term involving $M^{(0)}_z $ in Eq.~(\ref{eq:Mz_2nd_inte1st}) cancels out the second and third terms in Eq.~(\ref{eq:Mz_2nd}).  
Hence, we obtain
\begin{align}
M_z^{(2)}(t) 
= \frac{1}{\hbar^2 } \int^{t}_{\! -\infty} \! dt_1 \int^{t}_{\! - \infty} \! dt_2 \sum_{\zeta=\pm} \sum_{m} \zeta\braket{\psi_0|  \hat{\mathcal{V}}_{\rm I}(t_1) | \psi_m^{\zeta}} &
\notag \\
\times \braket{\psi_m^{\zeta} | \hat{\mathcal{V}}_{\rm I}(t_2) |\psi_0} & . 
\label{eq:Mz2nd_MatEl}
\end{align}
Combining the relations  
\begin{align}
&\braket{\psi_0|  \hat{S}^{\nu}| \psi_m^{\pm}} \braket{\psi_m^{\pm} | \hat{S}^{\nu} |\psi_0} 
= \frac{1}{4} |\braket{\psi_m^{\pm} | \hat{S}^{\pm}|\psi_0} |^2, 
\\
&\braket{\psi_0|  \hat{S}^{x}| \psi_m^{\pm}} \braket{\psi_m^{\pm} | \hat{S}^{y} |\psi_0} 
= \pm \frac{1}{4i} |\braket{\psi_m^{\pm} | \hat{S}^{\pm}|\psi_0} |^2, 
\end{align}
where $\nu = x,y$, we find 
\begin{align}
&\braket{\psi_0|  \hat{\mathcal{V}}_{\rm I}(t_1) | \psi_m^{\zeta}} \braket{\psi_m^{\zeta} | \hat{\mathcal{V}}_{\rm I}(t_2) |\psi_0}  
\notag \\
&= \frac{\hbar^2 \gamma^2}{4} |\braket{\psi_m^{\zeta} | \hat{S}^{\zeta}|\psi_0}|^2 
e^{-i(\omega_m^{\zeta}-\omega_0)(t_1-t_2)}  \bm{B}(t_1)\cdot \bm{B}(t_2)
\notag \\
&+ \frac{\hbar^2\gamma^2}{4i} \zeta |\braket{\psi_m^{\zeta} | \hat{S}^{\zeta}|\psi_0}|^2 
e^{-i(\omega_m^{\zeta}-\omega_0)(t_1-t_2)} \left[ \bm{B}(t_1) \! \times \! \bm{B}(t_2) \right]_z . 
\label{eq:MatEl}
\end{align}
Here, $\hbar \omega_0$ is the ground-state energy of $\ket{\psi_{0}}$ and $\hbar \omega^{\zeta}_m$  ($\zeta = \pm$) is the eigenenergy of $\ket{\psi_m^{\pm}}$. 
Then, applying Eq.~(\ref{eq:MatEl}) to Eq.~(\ref{eq:Mz2nd_MatEl}), we obtain Eq.~(\ref{eq:Mzt_all}). 

While the above formulas are the results at zero temperature, we may obtain the corresponding formulas at nonzero temperature by replacing $\braket{\psi_0|\cdots|\psi_0}$ with $1/Z \sum_n e^{-\beta \hbar \omega_n} \braket{\psi_n|\cdots|\psi_n} $, where $\beta$ and $Z$ are the inverse temperature and the partition function, respectively.

\section{Magnetization in a rotating frame} \label{appRF}

In this appendix, we consider magnetization in a rotating frame. 
In this frame, we can discuss the magnetization under the circularly polarized field 
as the dynamics described by the static Hamiltonian with an effective magnetic field~\cite{takayoshi2014_1,takayoshi2014_2}. 

With respect to the original Schr{\" o}dinger equation $[ i\hbar \frac{d}{dt} - \hat{\mathcal{H}}(t) ] \ket{\Psi(t)}=0$, a state given by a unitary transformation $\ket{\Psi ' (t)} = \hat{U}(t) \ket{\Psi(t)}$ satisfies~\cite{takayoshi2014_1}
\begin{align}
\hat{U}(t) \left[ i\hbar \frac{d}{dt} - \hat{\mathcal{H}}(t) \right] U(t)^{\dag} \ket{\Psi'(t)}=0.
\end{align}
Here, assuming the U(1) spin rotational symmetry around the $z$ axis in the Hamiltonian $\hat{\mathcal{H}}_0$, we apply 
\begin{align}
\hat{U}(t) = e^{i \xi \Omega \hat{S}^z t}, 
\end{align}
where $\xi = \pm 1$ denotes clock/anticlockwise rotation. 
Then, 
\begin{align}
\hat{U}(t)  i\hbar \frac{d}{dt} \hat{U}(t)^{\dag} = i\hbar \frac{d}{dt}+ \xi \hbar \Omega \hat{S}^z
\end{align}
and
\begin{align}
\hat{U}(t)  \hat{\mathcal{H}}(t) \hat{U}(t)^{\dag} = \hat{\mathcal{H}}_{0} 
 - \hbar \gamma  B^x(t) \left[ \hat{S}^x \cos \Omega t - \xi \hat{S}^y \sin\Omega t \right] &
\notag  \\
 - \hbar \gamma  B^y(t) \left[ \hat{S}^y \cos \Omega t + \xi \hat{S}^x \sin \Omega t  \right] &.
\end{align}
Hence, for the time-dependent equation 
\begin{align}
\left[ i\hbar \frac{d}{dt}- \hat{\mathcal{H}}'(t) \right] \ket{\Psi'(t)}=0, 
\end{align}
we find 
\begin{align}
 \hat{\mathcal{H}}'(t) 
&= \hat{\mathcal{H}}_{0} 
- \hbar \gamma \left[ B^x(t) \cos \Omega t + \xi B^y(t)\sin\Omega t \right] \hat{S}^x
\notag  \\
&- \hbar \gamma \left[ B^y(t) \cos \Omega t - \xi B^x(t)\sin\Omega t \right] \hat{S}^y
-  \xi \hbar \Omega \hat{S}^z. 
\end{align}

Here, we consider the case under the circularly polarized field. 
When frame rotation corresponds to the helicity of the magnetic field as $\bm{B}(t)= B_{\rm C} \left( \cos \Omega t , \xi \sin \Omega t \right)$, we obtain the static Hamiltonian 
\begin{align}
 \hat{\mathcal{H}}'_{\rm C} 
&= \hat{\mathcal{H}}_{0} - \hbar \gamma B_{\rm C} \hat{S}^x -  \xi \hbar \Omega \hat{S}^z 
\end{align}
for $[ i\hbar \frac{d}{dt} - \hat{\mathcal{H}}'_{\rm C} ] \ket{\Psi'(t)}=0$~\cite{takayoshi2014_1,takayoshi2014_2}. 
This static Hamiltonian in the rotating frame indicates that the frequency $\Omega$ gives the effective Zeeman term $ -  \xi \hbar \Omega \hat{S}^z $ and the magnetization direction depends on the helicity ($\xi$) of the circularly polarized field. 
Since $[\hat{\mathcal{H}}_{0}  - \xi \hbar \Omega \hat{S}^z,\hat{S}^z]=0$, the effective Zeeman term $ -  \xi \hbar \Omega \hat{S}^z $ itself cannot change the magnetization $M_z = \braket{\hat{S}^{z}}$ from the ground state $\ket{\psi_0}$ of $\hat{\mathcal{H}}_0$. 
However, the perturbation due to $- \hbar \gamma B_{\rm C} \hat{S}^x$ breaks the conservation of $S^z$ and can modify the magnetization in anisotropic magnets~\cite{takayoshi2014_1,takayoshi2014_2}. 
In this picture, when $M^{(0)}_z = 0$ at $t=-\infty$ in $\ket{\psi_0}$, the magnetization at the second order is given by 
\begin{align}
M_z^{(2)}(t) \! = \! \gamma^2 B^2_{\rm C} \! \int^{t}_{- \infty} \!\!\! dt_1 \int^{t}_{-\infty} \!\!\! dt_2 \braket{\psi_0|   \hat{S}^{x \prime}_{\rm I} \! (t_1)  \hat{S}_{\rm I}^{z \prime} \! (t)  \hat{S}^{x \prime}_{\rm I} \! (t_2)  |\psi_0} ,
\end{align}
where $\hat{\mathcal{O}}'_{\rm I} (t) = e^{\frac{i}{\hbar} (\hat{\mathcal{H}}_{0}  - \xi \hbar \Omega \hat{S}^z ) t} \hat{\mathcal{O}} e^{-\frac{i}{\hbar} (\hat{\mathcal{H}}_{0}  - \xi \hbar \Omega \hat{S}^z ) t} $ with respect to $\ket{\Psi'(t)} = e^{ -\frac{i}{\hbar} (\hat{\mathcal{H}}_{0}  - \xi \hbar \Omega \hat{S}^z ) t} \ket{\Psi'_{\rm I}(t)}$. 
Assuming $|\braket{\psi^+_m | \hat{S}^+|\psi_0}|^2  =  |\braket{\psi^-_m | \hat{S}^-|\psi_0}|^2$ ($\omega_m^+ = \omega_m^-$), we finally obtain 
\begin{align}
\frac{dM_z^{(2)}}{dt}
= \xi \frac{ \pi \gamma^2 B^2_{\rm C}}{2} \sum_{m} |\braket{ \psi_m |  \hat{S}^{-} |\psi_0} |^2 
\bigl[ \delta(  \Omega -\omega_m +\omega_0 ) &
\notag \\
- \delta(  \Omega +\omega_m -\omega_0 ) & \bigr] . 
\end{align}
This is consistent with Eq.~(\ref{eq:Tz_2nd}) since $i [\bm{B}(\Omega) \times \bm{B}(-\Omega)]_z = \xi B^2_{\rm C}/2$.

\section{Time evolution from $t=0$}  \label{appTP0}

In the above derivations, we assumed adiabatic switching from $t=-\infty$. 
Here, for real-time numerical simulations, we derive the formula when the magnetic field is switched at $t=0$.  
When $|\braket{\psi^+_m | \hat{S}^+|\psi_0}|^2  =  |\braket{\psi^-_m | \hat{S}^-|\psi_0}|^2$ and $\omega_m^+ = \omega_m^-$ are satisfied, the time-dependent magnetization in Eq.~(\ref{eq:Mzt_all}) under a monochromatic field  $\bm{B}(t) = \theta(t) \bm{B}(\Omega) e^{-i\Omega t} + {\rm c.c.}$ is given by 
\begin{align}
M_z^{(2)}(t) 
 = 2i\gamma^2 \sum_{m}  |\braket{\psi_m | \hat{S}^- |\psi_0}|^2 \! 
\Biggl[  
   \frac{\sin^2[( \Omega \!-\! \omega_m \!+\! \omega_0) t/2]}{(\Omega -\omega_m+\omega_0)^2}   
\notag    \\
-  \frac{\sin^2[( \Omega \!+\! \omega_m \!-\! \omega_0) t/2]}{(\Omega +\omega_m-\omega_0)^2}   
\Biggr]
 \left[ \bm{B}(\Omega) \! \times \! \bm{B}(-\Omega)\right]_z   ,
\end{align}
where $\ket{\psi^{\pm}_m}$ and $ \omega_m^{\pm}$ are denoted by $\ket{\psi_m}$ and $ \omega_m$.  
$M_z^{(2)}(t) \propto t^4$ at $t\sim 0$. 
On the other hand, in the limit $t\rightarrow \infty$, we find 
\begin{align}
M_z^{(2)}(t) 
 = i t \pi \gamma^2 & \sum_{m}   |\braket{\psi_m | \hat{S}^- |\psi_0}|^2
\Bigl[  
    \delta\left( \Omega - \omega_m+\omega_0 \right) 
    \notag \\
&\;\; -  \delta\left( \Omega + \omega_m-\omega_0\right) 
\Bigr]
 \left[ \bm{B}(\Omega) \! \times \! \bm{B}(-\Omega)\right]_z  .
\end{align} 
Hence, $M_z^{(2)}(t \rightarrow \infty) \propto t$ and its time derivative is consistent with Eq.~(\ref{eq:Tz_2nd}).


\bibliography{References}

\end{document}